# A Simulation of Lyα Absorption Forests in Liner Approximation of Cold and Cold plus Hot Dark Matter Models

Hongguang Bi[1*], Jian Ge[2], and Li-Zhi Fang[1]

[1] Department of Physics, University of Arizona, Tucson, Arizona 85721

[2] Steward Observatory and Physics Department, University of Arizona, Tucson, Arizona 85721

*All correspondence to: L.Z. Fang*

# Abstract

Our goal in this paper is to test some popular dark matter models by Ly$\alpha$ forest in QSO spectra. Recent observations of the size and velocity of Ly$\alpha$ forest clouds have indicated that the Ly$\alpha$ absorption is probably not given by collapsed objects, but pre-collapsed regions in the baryonic density field. Therefore, a linear approximation description would be able to provide valuable information. We developed a technique to simulate Ly$\alpha$ forest as the absorption of such pre-collapsed regions under linear approximation regime. The simulated Ly$\alpha$ forests in models of the standard cold dark matter (SCDM), the cold plus hot dark matter (CHDM), and the low-density flat cold dark matter (LCDM) have been confronted with observational features, including 1) the number density of Ly$\alpha$ lines and its dependencies on redshift and equivalent width; 2) the distribution of equivalent widths and its redshift dependence; 3) clustering; and 4) the Gunn-Peterson effect. The "standard" CHDM model, i.e. 60% cold and 30% hot dark matters and 10% baryons, is found to be difficult to pass the Ly$\alpha$ forest test, probably because it produces structures too late and favors to form structures on large scales instead of small scale objects like Ly$\alpha$ clouds. Within a reasonable range of $J_\nu$, the UV background radiation at high redshift, and $\delta_{th}$, the threshold of the onset of gravitational collapse of the baryonic matter, the LCDM model is consistent with observational data in all above-mentioned four aspects. The model of SCDM can also fit with observation, but it requires a smaller $J_\nu$ and a higher $\delta_{th}$. This suggests that whether a significant part of the Ly$\alpha$ forest lines is located in the halos of collapsed objects would be crucial to the success of SCDM.

## 1. Introduction

Ly$\alpha$ absorption line systems, shortward of Ly$\alpha$ emission in QSO spectra, indicate intervening absorbers with neutral hydrogen column densities ranging from about $10^{13}$ to $10^{22}$ cm$^{-2}$. The absorbers with low column densities, *e.g.* from $10^{13}$ to $10^{17}$ cm$^{-2}$, are usually called Ly$\alpha$ forest or Ly$\alpha$ clouds. In terms of metallicity, clustering, number density and redshift evolution, the Ly$\alpha$ clouds demonstrate very different properties from higher column density absorption systems. These differences probably come from the difference in their origins. It is generally thought that the high column density absorption systems are most likely due to halos or disks of foreground galaxies, while the low column density absorbers are some kind of less clustered clouds consisting of photoionized intergalactic gas (*e.g.* Blades, Turnshek & Norman 1988; Wolfe 1991; Bajtlik 1992).

Various models for the Ly$\alpha$ forest have been proposed, including pressure-confined clouds in hot intergalactic medium (Sargent et al. 1980; Ostriker 1988), self-gravitating objects (Bahcall & Salpeter 1965; Black 1981), minihalos in cold dark matter (CDM) scenario (Rees 1986; Murakami & Ikeuchi 1993), low mass objects formed from the short-wave tail in the power spectrum of the density perturbations in the CDM model (Bond, Szalay & Silk (1988). These models generally assumed that the Ly$\alpha$ forest absorber is very well developed object, such as static self-gravitating cloud, virialized baryonic cloud in potential well of dark matter, and stably-confined gas in pressure equilibrium. However, recent measurements have found that the size of the Ly$\alpha$ clouds at high redshift is unexpected as large as 100 - 200 h$^{-1}$ kpc (where h is the Hubble constant in unit of 100 km s$^{-1}$ Mpc$^{-1}$), and their velocity dispersion is unexpected as low as $\sim$ 100 km s$^{-1}$ (Bechtold et al. 1994, Dinshaw et al. 1995). These results cannot matched with the pictures of pressure equilibrium and virialization. For instance, if the Ly$\alpha$ clouds with such large size are well gravitational confined, the Press-Schechter theory show that their column density should be equal to or larger than $10^{17}$ cm$^{-2}$ (Mo, Miralda-Escudé & Rees 1993). Clouds with large size and low column densities are not completely gravity-confined.

Therefore, the Ly$\alpha$ clouds are probably neither virialized nor completely gravity-confined, but given by pre-collapsed areas in the baryonic density field. This suggests that Ly$\alpha$ clouds would still be remaining in the stage of evolution that non-linear process is less important. Therefore, linear approximation would be able, at least partially, to describe the features of Ly$\alpha$ absorption clouds. This might be the reason that the linear approximation simulation done by Bi, Börner & Chu (1992 hereafter BBC; Bi 1993 hereafter B93) successfully realized some statistical properties of Ly$\alpha$ absorption spectrum. Within a certain range of model parameters they found that the simulated spectrum is in a good agreement with the observed distribution of line equivalent width in the range $0.1\text{Å} \leq W \leq 0.5\text{Å}$. The two-point correlation function of the synthetical lines is also consistent with the observation.

In this paper, we will develop the technique of simulating Ly$\alpha$ absorption in QSO spectra in the linear approximation regime. Our goal is to use Ly$\alpha$ forest as a cosmological test to discriminate among some popular dark matter models, including the standard cold dark matter (SCDM), the cold plus hot dark matter (CHDM), and the low-density flat cold dark matter (LCDM). As we knew, objects at higher redshift can serve a very promising test to discriminate among models, which have been found in good agreement with the observational data of low redshift galaxy clustering and of the microwave background radiation (Jing & Fang 1994; Ma & Bertschinger 1994). One can expect that the distribution and evolution of Ly$\alpha$ clouds on a redshift range from 2 to 4 would also be able to play such important role. However, up to now, it is impossible to realize samples of $\sim$ 100 Kpc objects in so large redshift range by either hydrodynamical or N-body simulation. This is also a motivation of developing the simulation of Ly$\alpha$ forest in linear approximation.

The linear approximation will lead to uncertainties in the confrontation between observational and simulated samples, especially when clouds have density contrast larger than 1. Nevertheless, linear approximation approach is worth: 1) one can find some results which are less dependent on the linear approximation, for instance in the cases that the variance of density perturbation is less than 1; 2) considering the features of the uncertainties given by linear approximation, one can find some instructive constraints to models being considered.

In §2, we describe the method of the simulation. Considering the shortages in the works of BBC and B93, we will improve the simulation in the following ways: 1) using the density perturbation spectra normalized by the temperature fluctuation in the cosmic background radiation (Smoot et al. 1992; Bennett et al. 1994); 2) using a linear threshold to identify the regions of collapse and then to remove these regions from the density fields. In §3, we give the simulated results, and conducted a confrontation between real and simulated samples in various aspects, including the number density and its redshift evolution, the equivalent width distribution, clustering, and the Gunn-Peterson effect. Discussion and conclusion are shown in §4 finally.

## 2. The Method

We consider a flat universe. The mass density parameters satisfy $\Omega + \lambda = 1$, where $\Omega = \sum_i \Omega_i$, $\Omega_j$ ($j = c, h, b$) are the density parameters for cold, hot dark matter and baryons, respectively, and $\lambda = \Lambda/3H^2$ with $\Lambda$ the cosmological constant and $H$ the Hubble constant. Because we are only interested in density perturbations at redshifts less than 5, the mass density of the cosmic background radiation field can be neglected. An extreme interval of the present baryonic density given by the Big Bang nucleosynthesis is $\Omega_b = (0.006 - 0.03)$ h$^{-2}$ (Copi, Schramm & Turner 1995). In our simulation, we will choose $\Omega_b = 0.1$. In order to easily compare our results with other studies of structure formation (e.g. Jing & Fang, 1994), the parameters at $z = 0$ are taken to be $h = 0.5$, $\Lambda = \Omega_h = 0$ for SCDM; $h = 0.5$, $\Lambda = 0$, $\Omega_h = 0.30$ and $\Omega_c + \Omega_b = 0.7$ for CHDM; and $h = 0.75$, $\Omega_c + \Omega_b = 0.3$, and $\lambda = 0.7$ for LCDM.

*2.1 Linear density and velocity perturbations of baryonic matter*

The difficulty to study the evolution of the baryonic matter at late universe ($z \leq 5$) is the lack of knowledge on thermodynamical or hydrodynamical processes of the component. Since the first generation of luminous objects formed, the baryonic gas was strongly affected by energy and mass exchanges from the objects via processes such as heating, cooling, shocking, re-ionization, stellar winds and accretions. The details of these processes have poorly been known because of many uncertain factors related to the formation and evolution of QSOs and galaxies, and to their injection of energy and mass into intergalactic space. However, observations have shown that the statistical distributions of the Ly$\alpha$ forest, say HI column density $N_{HI}$ and Doppler parameter $b$, have only a little change in the redshift range $2 < z < 4$. It strongly indicates that the Ly$\alpha$ clouds do possess their own cosmological properties. To study these properties except for the proximity effect, it is reasonable to consider only cosmologically averaged contribution of the energy exchanging processes, but not necessarily to consider the details of individual luminous sources. Among these averaged processes, the most important one is the photo-ionization of baryons by a mean background UV radiation, i.e. the Gunn-Peterson effect (Gunn & Peterson 1965). Therefore, if our interests are limited to the properties of Ly$\alpha$ absorbers on large scales, the baryonic component can be treated as a cosmic component weakly linked to individual luminous objects. Thus, the baryonic matter could be approximately described as a polytropic gas with the equation of state $p \propto \rho^\gamma$. We have then

$$v_s(t) \equiv \frac{dp}{d\rho}\Big|_0 = \left(\frac{\gamma k T_0(t)}{m_p \mu}\right)^{1/2} \propto t^{(1-\gamma)}, \tag{1}$$

where $m_p$ is the mass of proton, $v_s(t)$ and $T_0(t)$ are, respectively, the sound velocity and the mean temperature of the baryonic matter at time $t$. Since no evolution of the temperature of the Lyα forest has been found, one can assume that the gas is isothermal, i.e. $\gamma = 1$. We will take the temperature to be $T = 5 \times 10^4$K corresponding to the velocity dispersion $b = 29$ km/s (*e.g.* Chaffee et al. 1986; Carswell et al. 1987; Rauch et al. 1992). In this case, the molecular weight $\mu$ in eq.(1) is $\sim 0.64$.

In the linear evolutionary stage, a good approximate solution for the density contrast $\delta_b$ and the peculiar velocity $\mathbf{v}_b$ of the baryonic matter in $k$-space has been found to be (Fang et al. 1993)

$$\delta_b(\mathbf{k}, t) \approx \frac{\delta_{DM}(\mathbf{k}, t)}{1 + x_b^2 k^2} ,  \tag{2}$$

$$\mathbf{v_b}(\mathbf{k}, t) \approx \frac{\mathbf{v}_{DM}(\mathbf{k}, t)}{1 + x_b^2 k^2} ,  \tag{3}$$

where $\delta_{DM}$ ($\equiv \Omega_c \delta_c + \Omega_h \delta_h$) and $\mathbf{v}_{DM}$ are the density and velocity perturbations of the dark matter, respectively. The Jeans length $x_b$ of the matter is given by

$$x_b \equiv \frac{1}{H_0} \left[ \frac{2\gamma k T(t)}{3\Omega \mu m_p (1+z)} \right]^{1/2} . \tag{4}$$

Eq.(2) shows that the baryonic matter traces the dark matter distribution on scales larger than $x_b$, and smooths over structures on scales less than $x_b$. The goodness of the approximate eqs.(2) and (3) have been checked by numerical solutions in SCDM and CHDM, and the solutions were used to model hot ($kT \sim 1$ keV) component in the intergalactic medium (Fang et al. 1993). It was successfully applied to calculate the cosmic Mach number, the Comptonization parameter and the soft X-ray background radiation. Therefore, at the linear evolutionary stage the density distribution of the baryonic matter can be described as a Gaussian perturbation with the power spectrum $P_b(k) = P_{DM}(k)/(1 + x_b^2 k^2)^2$, where $P_{DM}(k)$ is the spectrum of the dark matter. In the following simulation $P_{DM}(k)$ will be chosen from 1) Davis et al. (1985) for the SCDM model; 2) Klypin et al. (1993) for CHDM (since their formula fails at $k \geq 50$ Mpc$^{-1}$, we extrapolate it by $\propto k^{-3}$); and 3) Efstathiou, Bond & White (1992) for LCDM. All the spectra are normalized by the COBE's quadrupole anisotropy $Q = 13\mu K$ (Smoot et al. 1992).

## 2.2 Distribution of Lyα absorbers

All neutral hydrogen located in the line-of-sight of a given distant QSO will contribute to the Lyα absorption. Therefore, one can model the QSO Lyα forest as the absorption of HI in the density field calculated under the approximation developed in last section.

As mentioned in §1, recent observations of the size and velocity of the Lyα clouds at high redshifts implied that Lyα clouds are probably not virialized, and not located in the potential well of collapsed objects. Therefore, to model Lyα clouds, we should remove the regions corresponding to the collapsed objects from the density field of baryonic matter. The collapsed regions at high redshift might be related to well developed objects, such as CIV and MgII metal systems and dampled Lyα absorption systems. It was shown in the Press-Schechter (PS) formalism (1974; also in the peaks formalism, Bardeens et al. 1986) that the collapsed structures can be one-to-one mapped from their linear correspondences, i.e. it can approximately be identified as regions, in which the mean *linear* density contrasts are exceeding a threshold $\delta_{th}$. We will use a similar threshold to remove the collapsed regions.

Practically, we first do a convolution of the one-dimensional density field by a spatial Gaussian filter on the Jeans scale $x_b/\sqrt{3}$, where the factor $\sqrt{3}$ comes from the projection of 3-D to 1-D. We then label collapsed regions on all the peaks where the maximal density contrasts are higher than $\delta_{th}$. As in the PS formalism, the parameter $\delta_{th}$ essentially is semi-empirical. In our case, the only restriction would be $\delta_{th} > 1.69$. To reduce the uncertainty given by $\delta_{th}$, we will be mainly interested in the results that do not sensitively depend on the choice of $\delta_{th}$.

Therefore, the spatial distribution of baryons involving the QSO Lyα forest can be calculated in the following way. To find out a one-dimensional distribution of the matter, we project all modes of fluctuations in 3-D $k$-space into a given 1-D direction. The one-dimension power spectra of the density and velocity perturbations are given by

$$P_{1b}(k) = 2\pi \int_k^\infty P_b(q)qdq , \tag{5}$$

and

$$P_{1v}(k) = 2\pi H^2(t)a^2(t)k^2 \int_k^\infty P_b(q)q^{-3}dq. \tag{6}$$

respectively, where $a(t)$ denotes the cosmic scaling factor normalized at the present time, i.e. $a(t_0) = 1$. Discrete one-dimensional sample at redshift $z$, $\delta(k, z)$, can then be realized from the Gaussian probabilities with variances given by Eq.(5). The corresponding density fluctuation $\delta(x, z)$ in $x$ space can be obtained by the Fast-Fourier transformation. The spatial distribution of baryons at redshift $z$ will finally be found by removing all collapsed regions from the original $\delta_b(x, z)$. In the following, we will use the same notation $\delta_b(x, z)$ for the new distribution. Obviously, $\delta_b(x, z)$ contains also diffuse gas.

If the abundance of hydrogen and helium is constant over the perturbed distribution, the number density of neutral hydrogen atoms is

$$n_{HI}(x, z) = fn_H = f[1 + \delta_b(x, z)]n_C\Omega_H(1+z)^3 , \tag{7}$$

where $n_H$ is the number density of hydrogen atoms, $n_C = \rho_C/m_p$, $\rho_C = 3H_0^2/8\pi G$ is the present critical density, and $\Omega_H = X\Omega_b^0$, where $X = 0.76$ being the cosmic abundance for hydrogen. The factor $f \equiv n_{HI}/n_H$ accounts for the neutral fraction. If the medium is in ionization balance, one has

$$f = \frac{\alpha(T)}{\alpha(T) + \Gamma + 4\pi J_c G_H/n_H} , \tag{8}$$

where $\alpha$, $\Gamma$, and $4\pi J_c G_H/n_H$ are the rates of radiative recombination, collisional ionization and photoionization, respectively. The parameters in eq.(8) are: $G_H = 2.54 \times 10^8$, and $J_\nu = J_c\nu_c/\nu$, where $\nu_c$ is the Lyman limit frequency and $J_c$ the external UV background flux at frequency $\nu_c$ (Black 1981). From the proximity effect, the Lyman limit background flux is estimated to be $J_{-21} = 0.3$ to $3.0$, where $J_{-21}$ is defined by $J_c \equiv J_{-21} \times 10^{-21}$ cm$^{-2}$

s$^{-1}$ Hz$^{-1}$ sr$^{-1}$ (e.g. Lu, Wolfe & Turnshek 1991; Bechtold 1994). Recently, $J_{-21}$ is found to be $\sim 0.3$ at redshift $z \sim 4.2$ (Williger et al. 1994), which is less than the estimates at $z \sim 2.5$ by a factor of 3 - 10. Therefore, the UV background probably declined from $z \sim 2$ to $z \sim 4$. In our simulation, we will consider several possible models for the background flux.

*2.3 Ly$\alpha$ forest absorption spectrum*

The profile of the Ly$\alpha$ absorption in the spectrum of a QSO at redshift $z_1$ is calculated by $\exp\{-\tau(\nu)\}$, where the optical depth $\tau$ is given by (Weinberg 1972; BBC)

$$\tau(\nu) = \int_{t_1}^{t_0} n_{HI}(t)\sigma_a(\frac{\nu}{a})dt \;, \tag{9}$$

where $t_0$ denotes the present time, $t_1$ the time corresponding to redshift $z_1$, $\nu$ the observed frequency, which is related to a redshift of $Z \equiv (\nu_a/\nu) - 1$, and $\nu_a$ is the Ly$\alpha$ frequency in rest. The absorption cross section $\sigma_a$ is given by

$$\sigma_a = \frac{I_a}{b\sqrt{\pi}}V(\alpha, \frac{\nu - \nu_a}{b\nu_a}) \;, \tag{10}$$

where $b = \sqrt{2kT/m_p}/c$ is velocity dispersion in the unit of $c$, $\alpha \equiv 2\pi e^2\nu_a/3m_ec^3b = 4.8548 \times 10^{-8}/b$, $I_a = 4.45 \times 10^{-18}$ cm$^{-2}$ and $V$ is the Voigt function. Using Eq.(7), $\tau(\nu)$ can be rewritten as

$$\tau(Z) = \frac{I_a}{\sqrt{\pi}}n_C\Omega_H(1+Z)^2\zeta(Z) \;, \tag{11}$$

where $\zeta(Z)$ given by

$$\zeta(Z) = \int_0^{z_1} dz f[1 + \delta_b(x(z), z)]\frac{1}{H_0\sqrt{\Omega(1+z)^3 + \lambda}}\frac{1}{b}V(\alpha, \frac{z-Z}{b(1+Z)}) \;, \tag{12}$$

and the function $x(z) = (1/H_0)\int_0^z[\Omega(1+z)^3 + \lambda]^{-1/2}dz$ is a normalized comoving coordinate. For absorption lines with width $W < 0.7\text{Å}$, one can safely replace the Voigt function by an exponential profile. We have then

$$\zeta(Z) = \int_0^{z_1} dz f[1 + \delta_b(x(z), z)]\frac{1}{H_0\sqrt{\Omega(1+z)^3 + \lambda}}\frac{1}{b}e^{-[(z-Z)/(1+Z)+v(z)]^2/b^2} \;, \tag{13}$$

$v(z)$ is the peculiar velocity in the direction of the line-of-sight. To synthesize an absorption spectrum over a large redshift range, say 1.5 to 4.1, we divide this range into 13 bins, each has $\Delta z = 0.2$ and is centered at $z_n = n \times 0.2 + 1.3$ with $n = 1$ to 13. For a given $z_n$, we can use the approximation $\delta_b(x(z), z) \sim \delta_b(x(z), z_n)$, because the exponential factor in eq.(13) restricts that the integrand has significant contribution to $\zeta(Z)$ only at $z \simeq Z$. For each $z_n$, we generate 20 or, sometimes, 100 samples. We take the pixel size in $z$ space to be about the same as observations, say, $0.0593\text{Å}$.

As examples, three simulated absorption spectra at redshift 3 in SCDM, LCDM and CHDM are shown, respectively, in Figs.1a, b and c. The observational instrumental point-spread-function 35 km s$^{-1}$ is assumed, and the Poisson and background noises are also added. Fig. 1a is for SCDM with $J_{-21} = 0.3$; Fig. 1b for LCDM, $J_{-21} = 1.0$, and Fig. 1c for CHDM, $J_{-21} = 0.3$. In all three spectra, $\delta_{th}$ is taken to be 1.69. The spectra of LCDM and SCDM, indeed, appear the features of Ly$\alpha$ forest. We can then identify the absorption lines from the spectrum by the same way as observation.

## 3. Result

### 3.1 Number density of Ly$\alpha$ clouds and its evolution

The $z$-dependence of the number density of Ly$\alpha$ absorption lines with rest equivalent width $W$ greater than a threshold $W_{th}$ is generally described by a power law

$$\frac{dN}{dz} = \left(\frac{dN}{dz}\right)_0 (1+z)^\gamma , \tag{14}$$

where $(dN/dz)_0$ is the number density extrapolated to zero redshift, and $\gamma$ the index of evolution. If the hydrogen distribution is comoving in a flat universe, the number density should be $(dN/dz) \propto (1+z)^2/[\Omega(1+z)^3 + \lambda]^{1/2}$. The deviation of $dN/dz$ from the comoving curve implies an evolution of the population of Ly$\alpha$ clouds. The index $\gamma$ is found to depend on $W_{th}$: the larger $W_{th}$, the higher $\gamma$ (Lu et al. 1991; Bechtold 1994). Instead of the phenomenological eq.(14), we directly use the observational data of $dN/dz$ as a function of $z$ in the following discussion. In this paper, we are limited to study high redshift Ly$\alpha$ forests ($z \geq 1.6$). Low redshift Ly$\alpha$ absorption forests showed, sometimes, different features from that at high redshifts and might be questionable to be described by a linear theory.

The simulated and observed $z$-dependence of $dN/dz$ are plotted in Figs.2, 4 and 6 for $W \geq 0.32\text{Å}$, and Figs.3, 5 and 7 for $W \geq 0.16\text{Å}$. The observational data are taken from Bechtold (1994). The UV background is assumed to be constant with flux $J_{-21} = 0.3$ (dotted), 1.0 (dashed) and 3.0 (long dashed). The threshold contrast $\delta_{th}$ is taken to be 1.5 (Figs. 2, 3), 4.5 (Figs. 4, 5) and 7.5 (Figs.6, 7).

Obviously, comparing with the PS formalism, $\delta_{th} \geq 3$ would be too large for identifying collapsed regions. However, simulations with larger values of $\delta_{th}$ is worth, because we try to find results which are less dependent on $\delta_{th}$. Moreover, we also try to test whether the less-collapsed perturbations are enough to explain Ly$\alpha$ absorption. If a model can only fit to the observation for $\delta_{th} \geq 4.5$, it probably implies that the less-collapsed clouds are not enough to explain Ly$\alpha$ absorption, and a part of the clouds should be located in collapsed regions. We should also keep in mind that in the PS formalism the smoothing scale can be chosen arbitrarily, but for the baryonic matter considered in our model the smoothing scale must be larger than the Jeans length since below the scale thermal pressure overcomes gravity. Therefore, if we smooth just at the Jeans scale, a threshold value might be slightly higher than that in PS formalism.

Figs.2c - 7c shows that the model of CHDM always predicts too few absorbers at high redshifts for all parameter ranges considered. Even for $J_{-21} = 0.3$ and $\delta_{th} = 7.5$ the number of the Ly$\alpha$ absorbers given by CHDM are still significantly less than the real data. In fact, the curves of $dN/dz$ in the CHDM model are almost independent of $\delta_{th}$. This shows that the deficit of absorption in the CHDM model is due to the late formation of structures. Because the range of $\delta_{th} \geq 1.5$ is located in the high density contrast tail of the perturbation, the thresholds $\delta_{th}$ remove only a very small fraction from the original $\delta_H$.

Therefore, the CHDM model is difficult to pass the test concerning subgalactic structures at high redshift. Actually, in the case of the CHDM model, the square of variance of the baryonic matter at the considered high redshift is only about 0.5, i.e. it is truly in the stage of linear evolution. Therefore, the failure of CHDM would not easily be overcame by non-linearity.

The results of SCDM sensitively depend on $\delta_{th}$. Figs. 2a - 7a showed that the better fittings are given by $\delta_{th} > 4$, i.e. less removal of the collapsed objects. It implies that in the SCDM model there are too much powers of the density perturbation on scales of few to tens Mpc, and baryons were already dragged into collapsed regions at very early time. Therefore, the baryons remaining in pre-collapsed regions are not enough to contribute to Ly$\alpha$ clouds. Therefore, the SCDM model requires a significant fraction of Ly$\alpha$ forest lines produced by gas component located in the regions of collapsed objects.

As has been known, comparing with the SCDM, the LCDM model has lower perturbation powers on scales of few kpc to about 10 Mpc, an overproduction of collapsed objects could be avoided. It will give enough pre-collapsed baryons to form the Ly$\alpha$ forest. On the other hand, unlike CHDM, LCDM predicts a significant development of clustering at $z > 1$, Ly$\alpha$ forest systems and other absorption systems could be formed early enough. As a result, $dN/dz$ in the LCDM model shows good fitting for all reasonable ranges of parameters (Figs. 2b - 7b). Both the $W_{th} = 0.32 \text{Å}$ and $0.16 \text{Å}$ curves of $dN/dz$ with $J_{-21} = 3$ in LCDM are consistent with the observations within 1 $\sigma$.

*3.2 Distribution of equivalent widths*

Within a given redshift range, the distribution of the rest frame equivalent width $W$ is given by

$$n(W) \propto e^{-W/W^*}, \tag{15}$$

where $W^*$ is found to be from about 0.23 Å to 0.35 Å (Sargent et al. 1980; Murdoch et al. 1986; Rauch et al. 1992; Weymann 1992). $W^*$ may also depend on redshift in the sense that the distribution steepens at lower redshift (Bechtold 1994).

Fig.8 shows the probability distributions of the rest equivalent widths in the $J_{-21} = 3$ and $\delta_{th} = 4.5$ LCDM model. The solid curve is the mean from the samples at $z = 2.6, 2.8, 3.0$ and $3.2$; the dotted curve from the samples at $z = 1.8, 2.0, 2.2$ and $2.4$. The theoretical distribution $n(W)$ has been normalized to the observation (Bechtold 1994) at $W = 0.32\text{Å}$; this is equal to normalize the total observed $W \geq 0.16\text{Å}$ lines at $z = 2.44$. The agreement between the model and the observation is rather good. This result cannot simply be explained as a less dependence of the width distribution on models. Actually, the width distribution is sensitive to models. For instance, we calculated distribution of the equivalent widths in the model CHDM with $J_{-21} = 0.3$ and $\delta_{th}$ (Fig.9). Both the $z > 2.5$ and $z < 2.5$ distributions are much steeper than the observational data. There is almost no $W \geq 0.6\text{Å}$ lines produced at $z < 2.5$. Therefore, the CHDM model is also uncomfortable in the test of the width distribution.

The $W$ distributions at individual redshifts 2.0, 2.4, 2.8, 3.2, 3.6 and 4.0 in $J_{21} = 3$ LCDM model are plotted in Fig.10. The dotted line in Fig.10 is given by eq.(15) with $W^* = 0.27\text{Å}$. A rapid evolution of the width distribution is found, especially at era after $z = 2.4$.

*3.3 Clustering properties*

Ly$\alpha$ clouds show an absence of two-point correlation in their spatial distribution. Some early samples were found to be excessive correlation at velocity differences less than $\sim 300$ km s$^{-1}$ (Webb 1987). Later samples usually detected no clustering at all. This fact has been used as a negative grade of the model of dark matter confinement, because it was argued that if Ly$\alpha$ clouds formed in the density field of dark matter, the spatial correlations of the clouds should be about the same as galaxies (e.g. Bajtlik 1992). However, it has already been shown in B93 that the simulated Ly$\alpha$ lines correlate weakly each other, even the density field is produced by a power law spectrum of the perturbations. Since collapsed regions have been removed, one can expect that the line-line correlations would be more weak.

The two point correlation functions of $W \geq 0.16 \text{\AA}$ lines for various redshifts in the LCDM model are plotted in Fig.11. In the entire redshift range from 2.0 to 4.0, no significant structures are detected. Therefore, the simulated spectra are, indeed, consistent with what observed, i.e. no cloud-cloud correlations exist on scales from 100 km s$^{-1}$ to 2000 km s$^{-1}$. It should be pointed out that the comparison cannot be taken seriously on scale less than 100 km s$^{-1}$ or 1 h$^{-1}$Mpc, because of the line-blending caused by the velocity dispersion and the point-spread-functions of FWHM$\sim$ 30 km/s. As a consequence, any clustering on that scales will essentially be removed. An anti-correlation is then to appear in Fig.11. It should also be pointed out that the conclusion of no line-line correlation make sense only statistically. It does not exclude rare events, such as possible voids (Dobrzycki & Bechtold 1991).

*3.4 The Gunn-Peterson test*

The simulated density contrast $1+\delta_H(x,z)$ contains hydrogen gas in both clumpy and diffuse forms. Therefore, the Gunn-Peterson optical depth, which came from the diffuse component, can directly be estimated from the minimums in $\tau(Z)$. No recognizable Gunn-Peterson depth can be found in the models of SCDM and LCDM. Assuming the signal-to-noise ratio S/N $\sim$ 29 for the continuum in the simulation samples, the upper limit of the depth is $\simeq 0.01$ from $z = 1.6$ to 4 in SCDM and LCDM.

However, for the model of CHDM with $J_{-21} = 0.3$, which is relatively the best one among all fluxes being considered, the Gunn-Peterson optical depth is found to be $\tau_{GP} \simeq 0.04$ at $z = 3.4$; 0.07 at $z = 3.6$; and 0.10 at $z = 4.0$ (Fig.12). From a measurement of $\tau_{GP}$ by spectrum of QSO 0000-263 ($z_{em} = 4.11$), we have $\tau_{GP} \leq 0.04$ at $z \approx 3.8$ (Webb et al 1992). A new limit, $\tau_{GP} \leq 0.02$ at $z \approx 4.3$, is obtained by Giallongo et al. (1994)

from QSO 1202-0725 ($z_{em} = 4.7$). Therefore, CHDM cannot pass the Gunn-Peterson test. This shows once again that the late structure formation in the CHDM model cannot be consistent with observations at high redshift.

## 4. Conclusions and Discussions

We simulated Ly$\alpha$ forests in QSO's spectra in linear regime, in which the Ly$\alpha$ clouds are modeled as the absorption of HI in less developed regions of the density field. Collapsed regions in the density field are identified, and then removed by a Press-Schecter-like criterion, i.e. a threshold of density contrast. The simulated samples have been compared with the observational data in four aspects: 1) the number density of Ly$\alpha$ lines and its dependencies on redshift and equivalent width; 2) the distribution of equivalent widths and its redshift dependence; 3) clustering; and 4) the Gunn-Peterson effect.

In spite of uncertainties in parameters of the UV background $J_{21}$ and the contrast threshold $\delta$, our simulation has shown that the Ly$\alpha$ forest can set effective constraints to models of large scale structure formation. The current observations have already given almost no room to the CHDM model even when the large uncertainties in the parameters are considered. In fact, one can see from Fig.1 that the simulated spectrum in the CHDM model fails to show the forest-like features. The square of variance of the baryonic matter in CHDM model is $\leq 0.5$ in redshift range $2 < z < 4$. Non-Linear evolution is negligible.

This result still hold for a larger normalization of the amplitude of the density perturbations. The two years of COBE-DMR observation (Bennett 1994) showed that the amplitude of the quadrupole anisotropy of cosmic background radiation might be as large as $Q_{rms-PS} = 18 \pm 1.5 \mu K$. Using this amplitude we repeated the simulation. The results for the CHDM model are plotted in Fig.13, which contains two sets of the curves $dN/dz$ vs. $(1+z)$, corresponding to the normalization of $Q_{rms-PS} = 18$ and $13 \mu K$, respectively. One can see from Fig.13 that the larger normalization do rise the number of Ly$\alpha$ clouds. However, there are still much less than observations.

This result is in a good agreement with other studies on larger scale structure formation in the CHDM model. The "standard" CHDM model of $\Omega_c/\Omega_h/\Omega_b = 0.6/0.3/0.1$ was already found to be lack of perturbation powers to form high redshift collapsed objects. The N-body simulation of CHDM at moderate redshift showed that a rapid formation of collapsed objects appeared only in the near past $z \simeq 0.5$ (Jing & Fang 1994). CHDM is also difficult in explaining the formation of damped Ly$\alpha$ systems (Klypin et al. 1994), and high-redshift halos (Ma & Bertschinger, 1994). Therefore, our results strengthened the conclusion that the "standard" CHDM model, i.e. 60% cold and 30% hot dark matters and 10% baryons, would be in problem to pass tests of high redshift objects.

When $J_{21} \leq 0.3$, all results of $dN/dz$ curves in LCDM model are larger than real data. Considering the linear approximation probably underestimate the number of clouds at high redshift, the simulated $dN/dz$ of LCDM would be a lower limit to the real number. When non-linear evolution is considered, the simulated result will be more larger than real data. Therefore, $J_{21} \leq 0.3$ should be a constraint to the model of LCDM.

The deficit of the $W \geq 0.32\text{Å}$ lines at $z \leq 2.5$ in $J_{21} = 3.0$ LCDM (Figs.2b, 4b and 6b), and the deficit of $W > 0.5\text{Å}$ objects in $z < 2.5$ width distribution of Fig.8 can be addressed on the uncertainties of parameters. If we replace the constant UV background by an evolutionary model, say the "turned on" model of QSOs at redshift $z = 5$ (Bajtlik, Duncan & Ostriker 1988; Bechtold 1994), the deficit disappeared (see, dot-dashed lines in Fig.14). Therefore, within a reasonable range of parameters, LCDM can reproduce all observational properties of the Ly$\alpha$ forest.

We should also pointed out the uncertainties in the current observations. First, the measurements of $dN/dz$, especially at $z \sim 4$, are still dispersive. Although the evolution of $dN/dz$ given by different groups (Lu et al. 1991; Press & Rybicki 1993; and Bechtold 1994) showed a common trend, their values of $dN/dz$ are differences by about 30%. Second, we have taken the velocity broadening parameter $b$ to be constant. This assumption is based on the observation that the distribution of $b$, well concentrated around 25~30 km s$^{-1}$, does not significantly correlate to column density $N_{HI}$ and to redshift. However, some observations showed the redshift-dependence of $b$ (*e.g.* Williger et al. 1994; Pettini et al. 1990).

Therefore, one cannot undoubtedly discriminate between SCDM and LCDM by the current simulation and real data. Nevertheless, a systematic difference between SCDM and LCDM can be seen significantly: good fitting of SCDM is generally given by higher threshold $\delta_{th}$ than in LCDM. In other word, the SCDM model needs an undercutting of collapsed regions than the LCDM, i.e. it appears to favors a significant part of the Ly$\alpha$ clouds to be related to halos of collapsed objects at high redshifts. The low redshift Ly$\alpha$ absorption lines have already been found to be partially produced by galactic halos (*e.g.* Mo & Morris 1994). This problem now is also important in high redshifts. Because the number of the collapsed regions identified by, say, PS formalism should be equal to or larger than the number of observed collapsed objects at high redshift, such as CIV or MgII metal systems in QSOs absorption spectrum, the occurrence of halo related Ly$\alpha$ forest absorptions might be estimated by the number density of these metallicity systems (Bi & Fang 1995).

We thank Dr. Y.P. Jing for useful discussions. JG thanks also Dr. J. Bechtold for her helps. HGB thanks World Lab for a fellowship. This work was partially support by the NSF INT-9301805.

## References

* World Lab Fellowship, Beijing Astronomical Observatory, Beijing, P.R. China

Bahcall, J.N. & Salpeter, E.E. 1965, ApJ, 142, 1677

Bajtlik, S. 1992, in The Environment and Evolution of Galaxies, ed. Shull, J. M. & Thronson, H. A. Jr. 191 (Klumer Academic Publisher)

Bajtlik, S., Duncan, R.C. & Ostriker, J.P. 1988, ApJ, 327, 570

Bardeen, J.M., Bond, J.R., Kaiser, N. & Szaley, A.S. 1986, ApJ, 304, 15

Bechtold, J. 1994, ApJSS, 91, 1

Bechtold, J., Crotts, P.S., Duncan, R.C. & Fang Y. 1994, ApJ, 437, L83

Bennett, C.L. et al. 1994, ApJ, 436, 423

Bi, H.G. 1993, ApJ, 405, 479 (B93)

Bi, H.G,. & Fang, L.Z. 1995, in preparation

Bi, H.G., Börner, G. & Chu, Y. 1992, A&A, 266, 1 (BBC)

Black, J.H. 1981, MNRAS, 1981, 197, 553

Blades, J.C., Turnshek, D.A. & Norman, C.A. 1988, in QSO Absorption Lines: Probing the Universe

Bond, J.R., Szaley, A.S. & Silk, J. 1988, ApJ, 324, 627

Carswell, R.F., Webb, J.K., Baldwin, J.A. & Atwood, B. 1987, ApJ, 319, 709

Chaffee, F.H., Foltz, C.R., Bechtold, J. & Weymann, R.J., 1986, ApJ, 116

Copi, C.J., Schramm, D.N. & Turner, M.S. 1995, Science, 267, 192

Davis, M., Efstathiou, G., Frenk, C.S. & White, S.D.M. 1985, ApJ, 292, 371

Dinshaw, N., Foltz, C.B., Impey, Weymann, R. & Morris, S.L. 1995, Nature, in press

Dobrzycki, A. & Bechtold, J. 1991, ApJ, 377, L69

Efstathiou, G., Bond, J.R. & White, S.D.M. 1992, MNRAS, 258, 1p

Fang, L.Z., Bi, H.G., Xiang, S.P & Börner, G. 1993, ApJ, 413, 477

Giallongo, E. et al, 1994, ApJ, 425, L1

# Figure Captions

**Figure 1** Examples of simulated spectra of the Ly$\alpha$ forest: (a) SCDM, $J_{-21} = 0.3$, where $J_{-21}$ is related to $J_c$ by $J_c \equiv J_{-21} \times 10^{-21}$ cm$^{-2}$ s$^{-1}$ Hz$^{-1}$ sr$^{-1}$; (b) LCDM, $J_{-21} = 1.0$; and (c) CHDM, $J_{-21} = 0.3$. In all three spectra, the threshold parameter $\delta_{th} = 1.69$.

**Figure 2** Number density $dN/dz$ of Ly$\alpha$ lines at $W_{th} = 0.32\text{Å}$ as a function of redshift. The threshold $\delta_{th} = 1.5$. The UV background is $J_{-21} = 0.3$ (dotted line); 1.0 (dashed line); and 3.0 (long dashed line). The open circles are data from Bechtold (1994). Models (a), (b), and (c) are for SCDM, LCDM, and CHDM, respectively.

**Figure 3** Number density $dN/dz$ of Ly$\alpha$ lines at $W_{th} = 0.16\text{Å}$ as a function of redshift. The threshold $\delta_{th} = 1.5$. The UV background is $J_{-21} = 0.3$ (dotted line); 1.0 (dashed line); and 3.0 (long dashed line). The open circles are data from Bechtold (1994). (a), (b), and (c) are for models of SCDM, LCDM, and CHDM, respectively.

**Figure 4** The same as Fig.2, but $\delta_{th} = 4.5$.

**Figure 5** The same as Fig.3, but $\delta_{th} = 4.5$.

**Figure 6** The same as Fig.2, but $\delta_{th} = 7.5$.

**Figure 7** The same as Fig.3, but $\delta_{th} = 7.5$.

**Figure 8** Distributions of rest equivalent widths in LCDM with $J_{-21} = 3.0$. Simulated results: low redshifts $1.8 \leq z < 2.5$ (dotted line); high redshifts $2.5 < z \leq 3.2$ (solid line). Observational data: $z < 2.5$ (open circle); $z > 2.5$ (solid circle). All distributions are normalized to the observed numbers at $0.32\text{Å}$.

**Figure 9** The same as Fig.8, but for the model of CHDM.

**Figure 10** Distributions of rest equivalent widths in the LCDM model with $J_{-21} = 3.0$ at redshifts 2.0, 2.4, 2.8, 3.2, 3.6 and 4.0. The dashed line is given by Eq.(15) with $W^* = 0.27 \text{Å}$.

**Figure 11** Two-point correlation functions of Ly$\alpha$ forest lines with $W \geq 0.16 \text{Å}$ in velocity space for model of LCDM with $J_{-21} = 3.0$. The error bars represent 1 $\sigma$.

**Figure 12** The Gunn-Peterson optical depth in the model of CHDM with $J_{-21} = 0.3$,

**Figure 13** Number density $dN/dz$ of Ly$\alpha$ lines at $W_{th} = 0.32$ and 0.16 Å as a function of redshift in CHDM model. The threshold $\delta_{th} = 4.5$. For dotted and dot-dashed lines, the UV background is $J_{-21} = 0.3$, and the normalization of the amplitude of initial perturbation is taken to be $Q_{rms-PS} = 18$ and $13 \mu K$, respectively. For dashed and solid lines, the UV background is $J_{-21} = 1.0$, and the normalization of the amplitude of initial perturbation is taken to be $Q_{rms-PS} = 18$ and $13 \mu K$, respectively. The open circles are data from Bechtold (1994).

**Figure 14** Number density $dN/dz$ and its $z$-dependence in the models of LCDM with $\delta_{th} = 4.5$. The UV background is taken to be a) constant $J_{-21} = 0.3$ (long dashed line), and b) the model of QSO "turn on" at $z = 5$ (dot-dashed line).

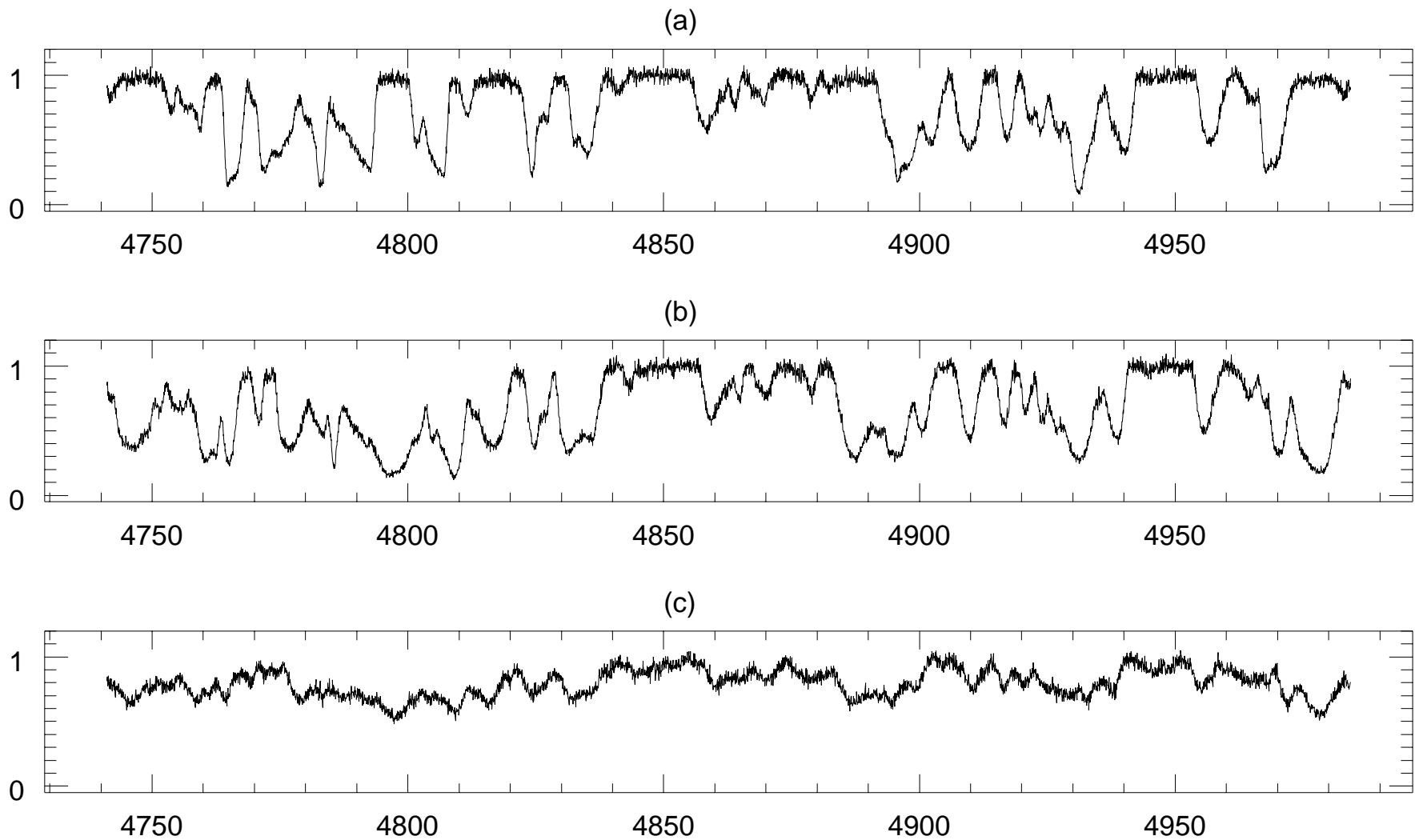

Fig. 1

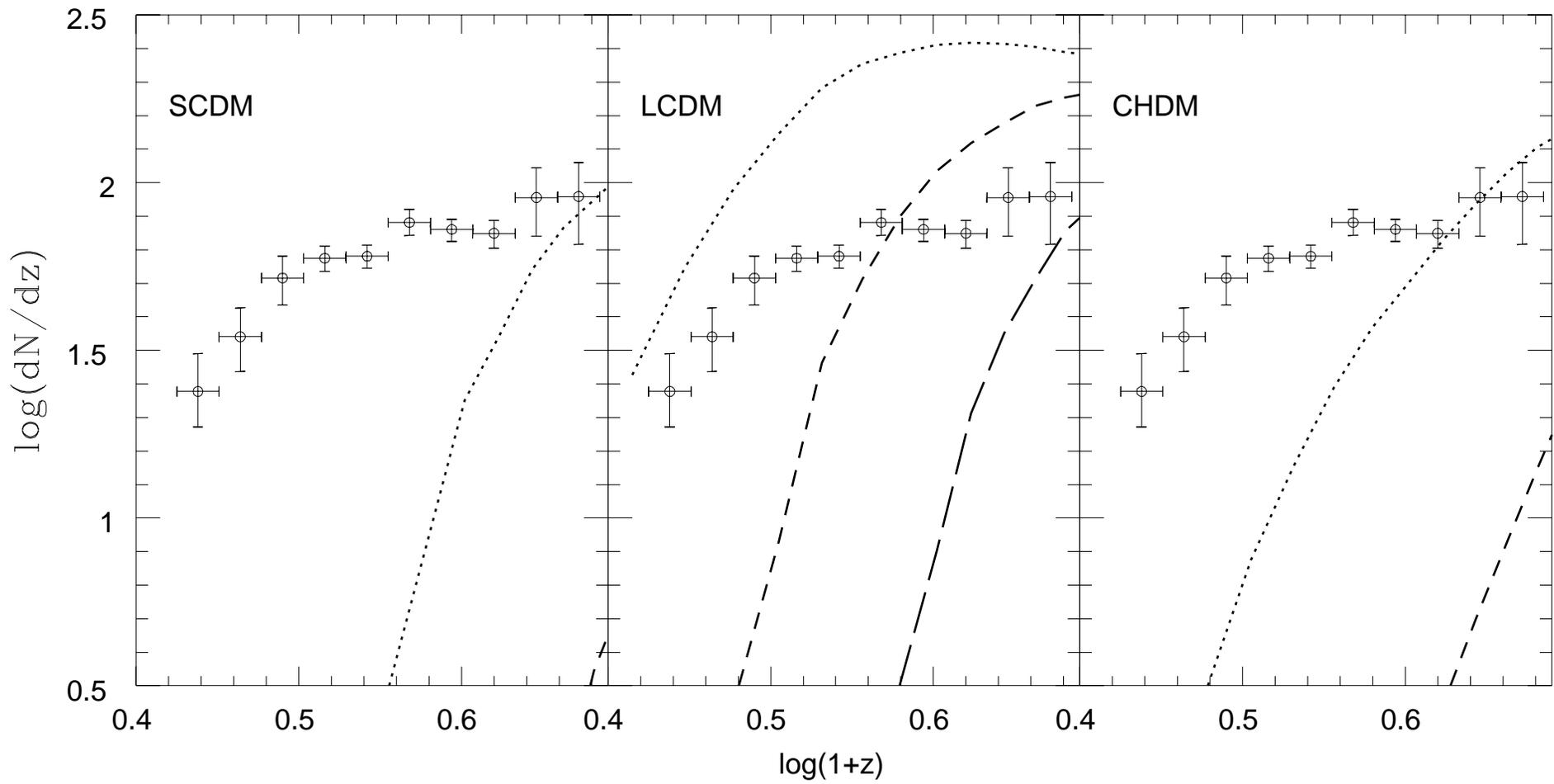

Fig. 2

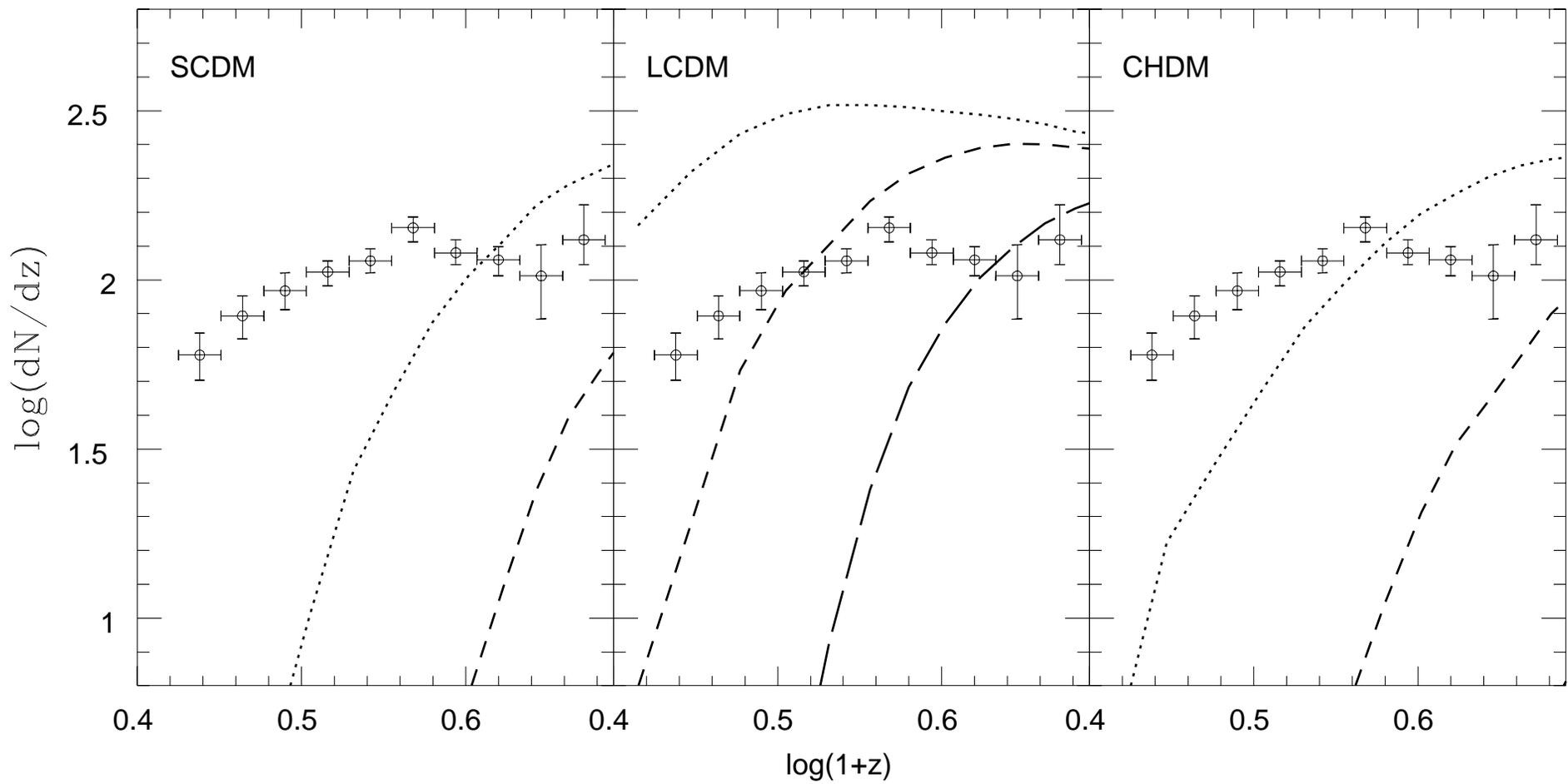

Fig. 3

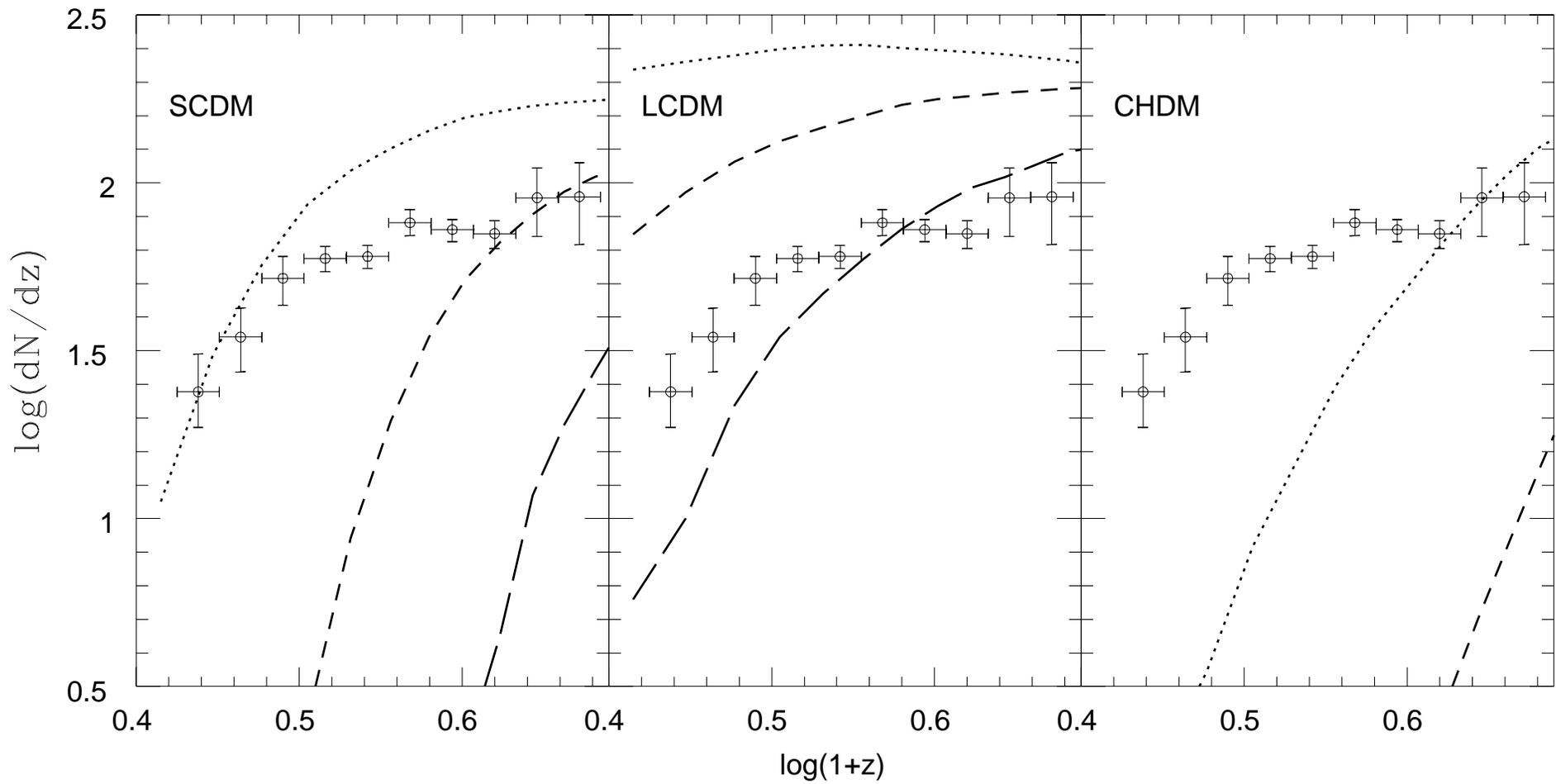

Fig. 4

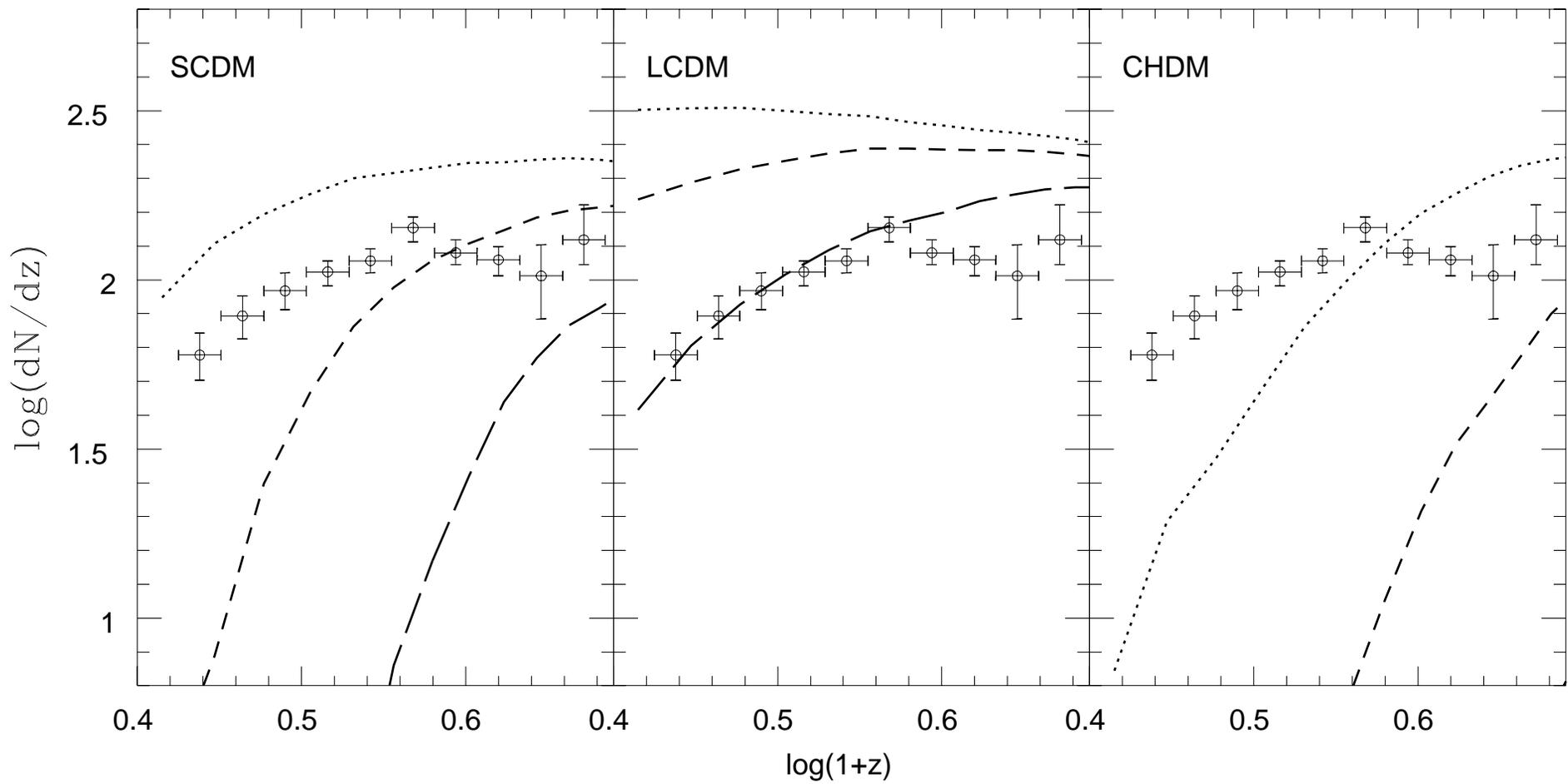

Fig. 5

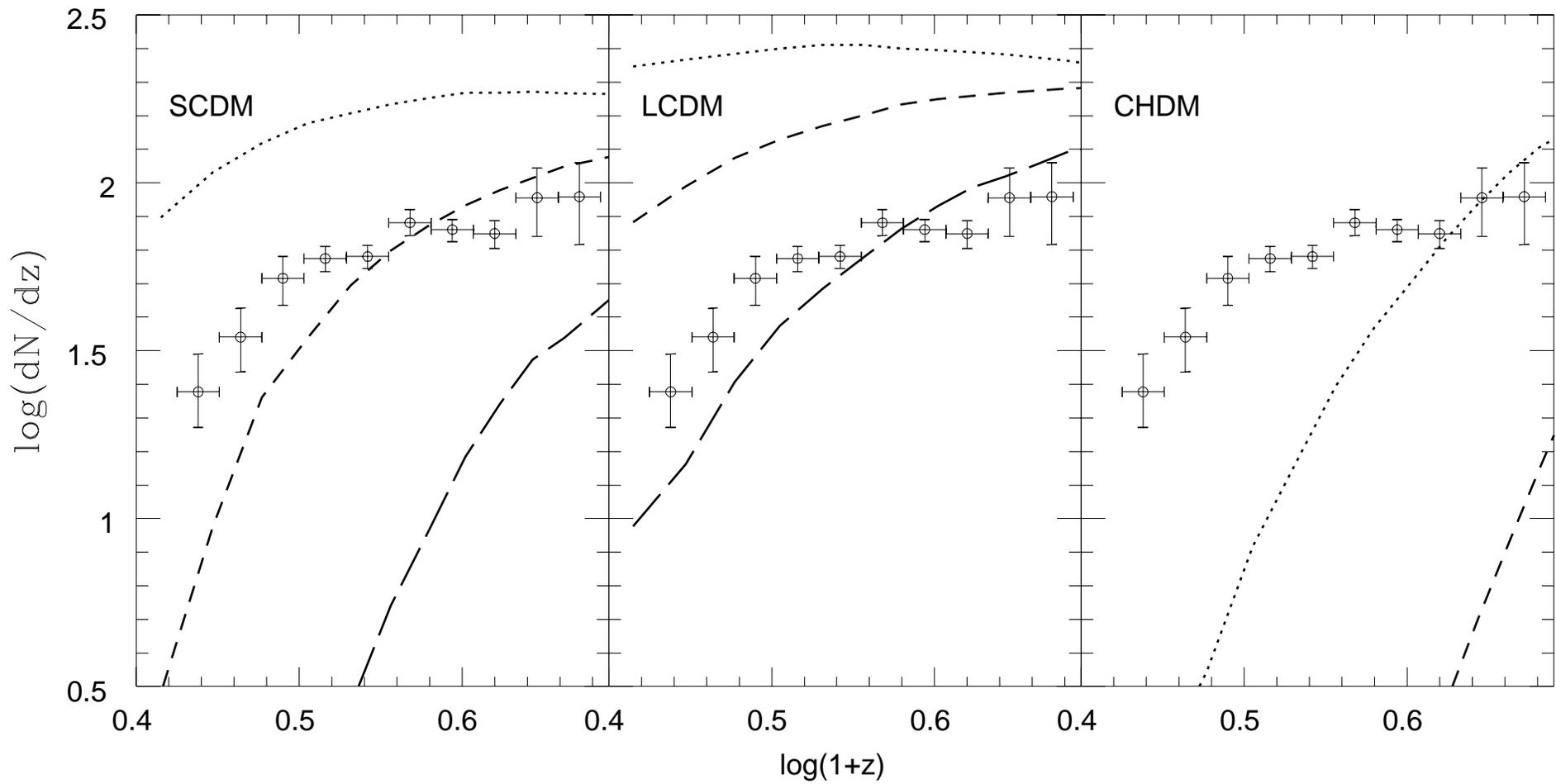

Fig. 6

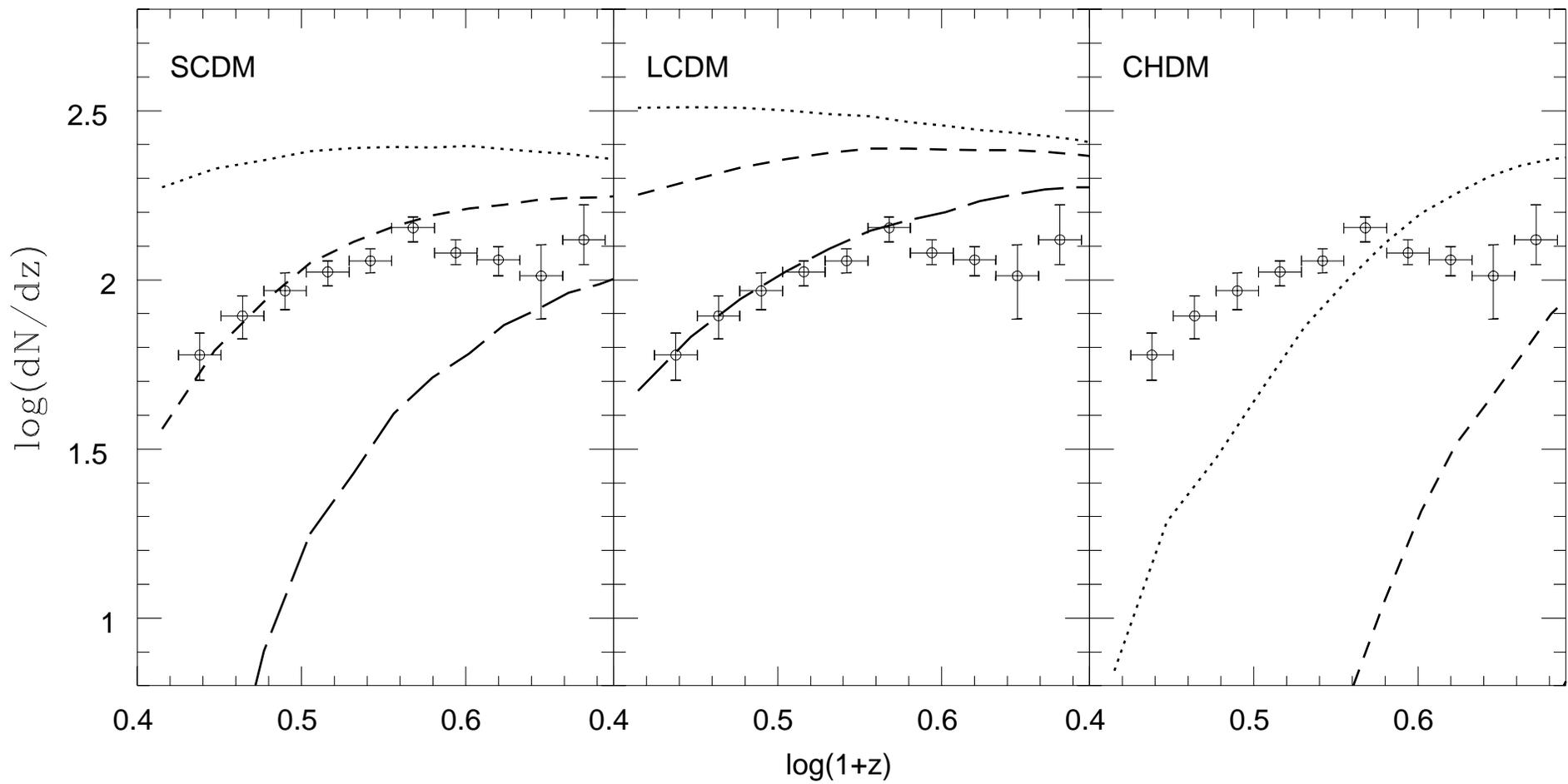

Fig. 7

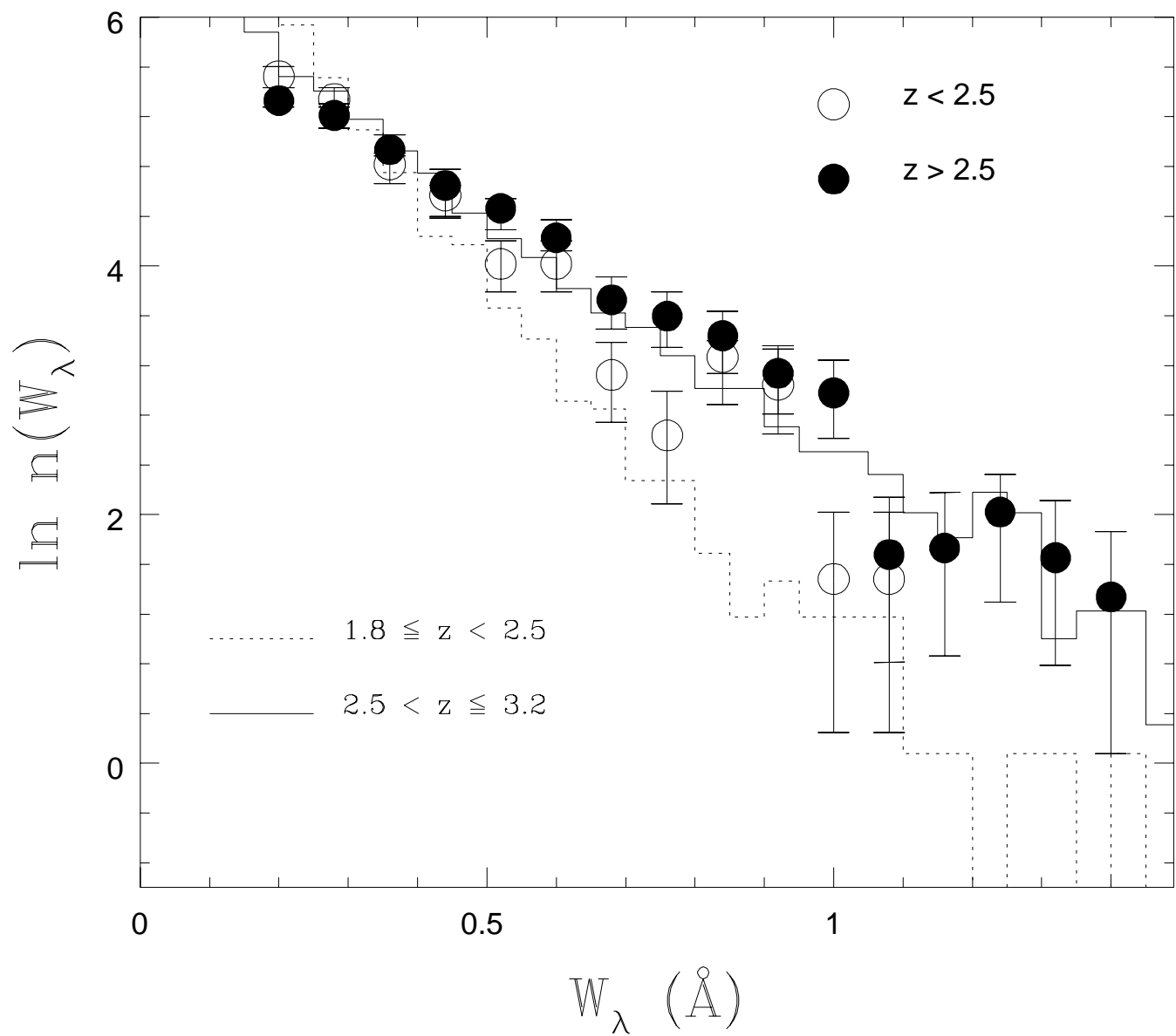

Fig. 8

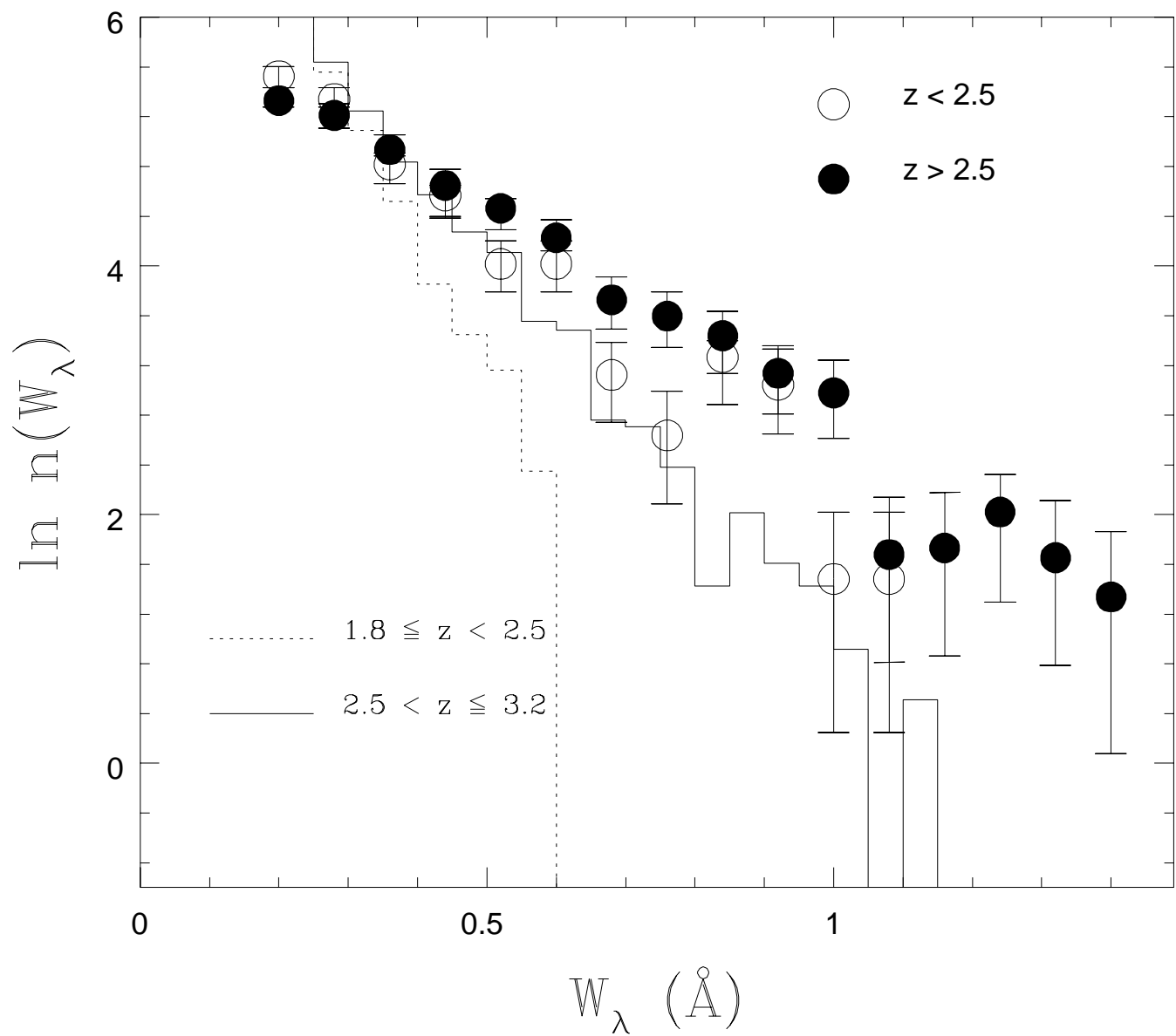

Fig. 9

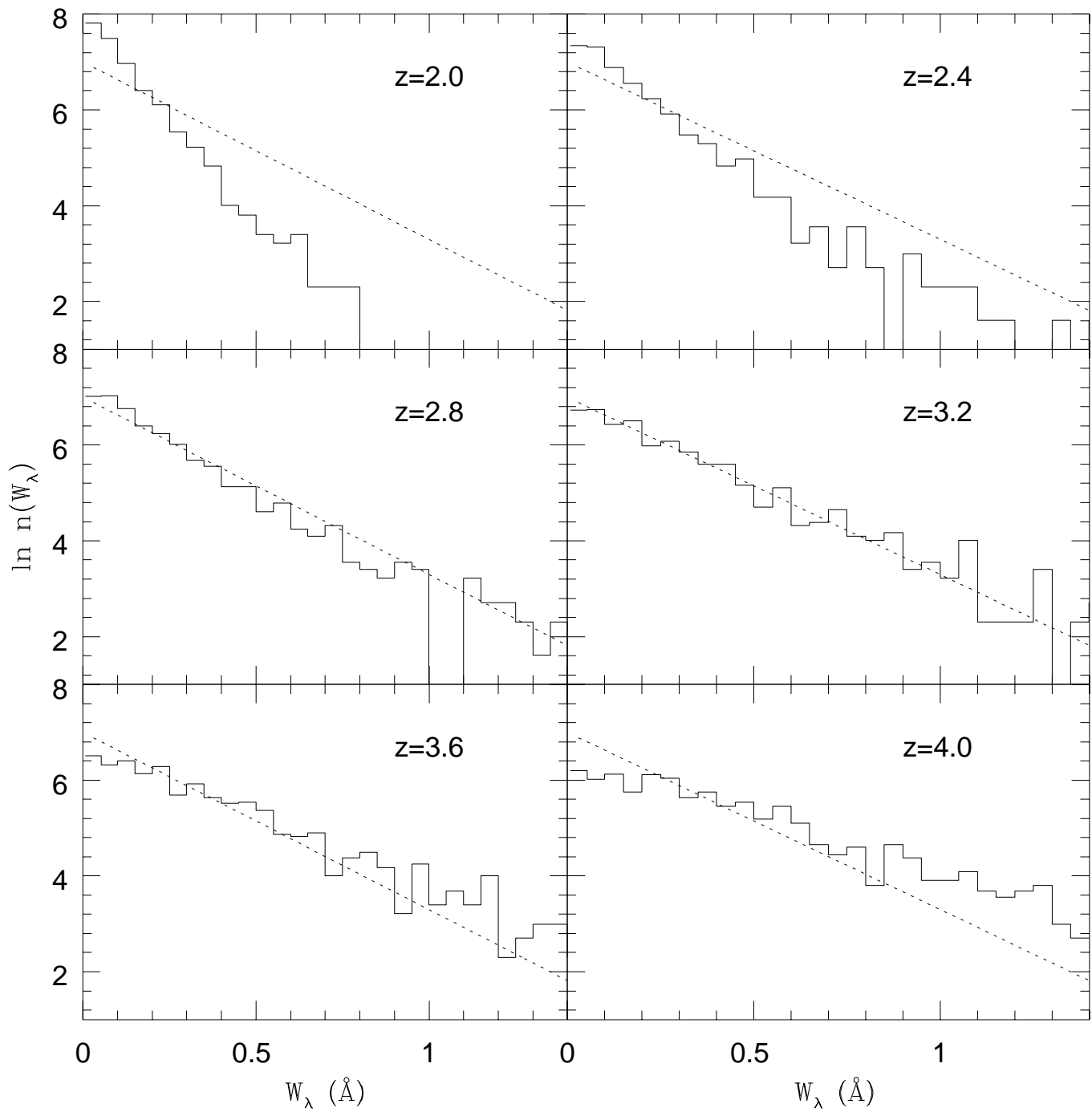

Fig. 10

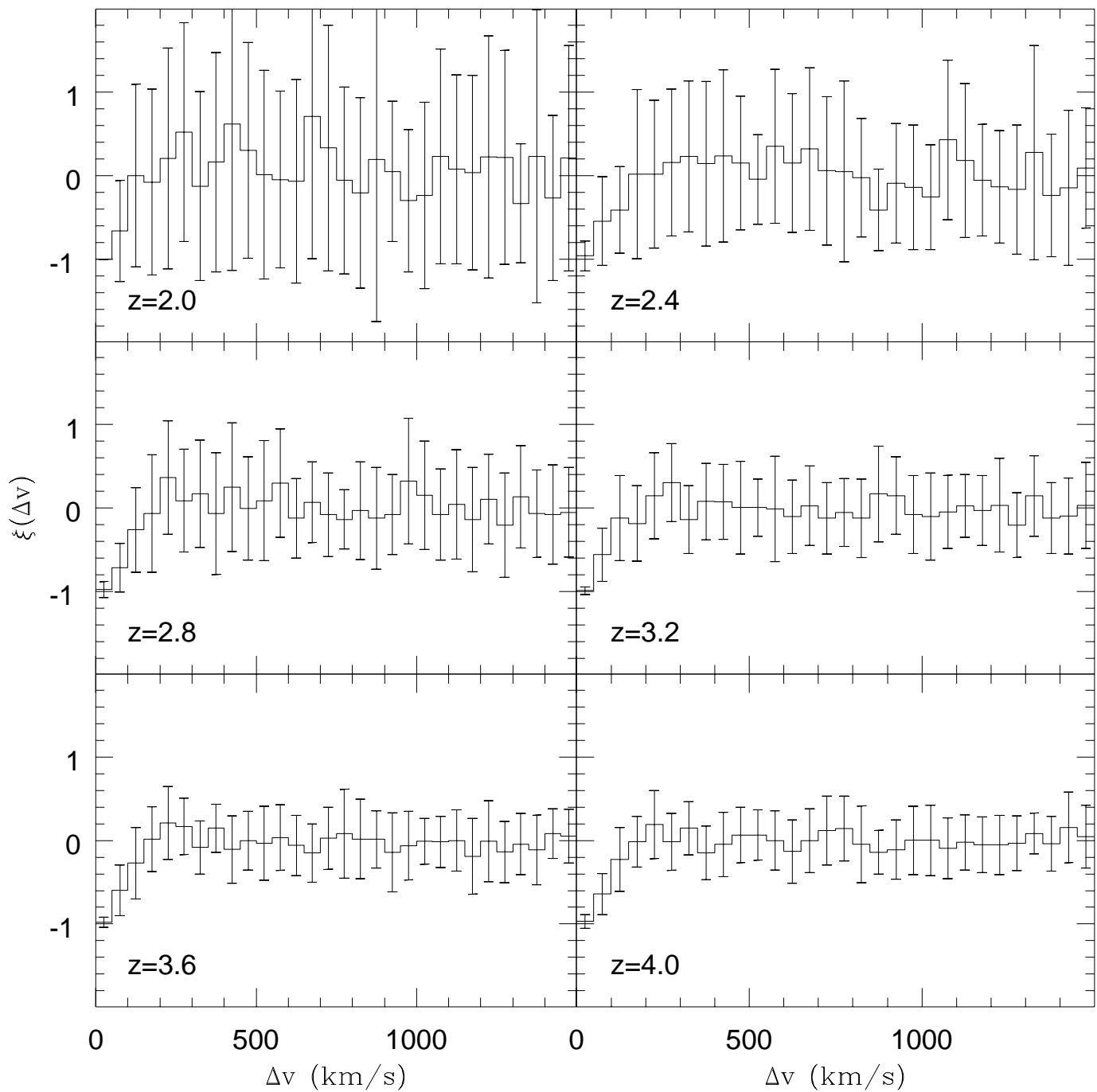

Fig. 11

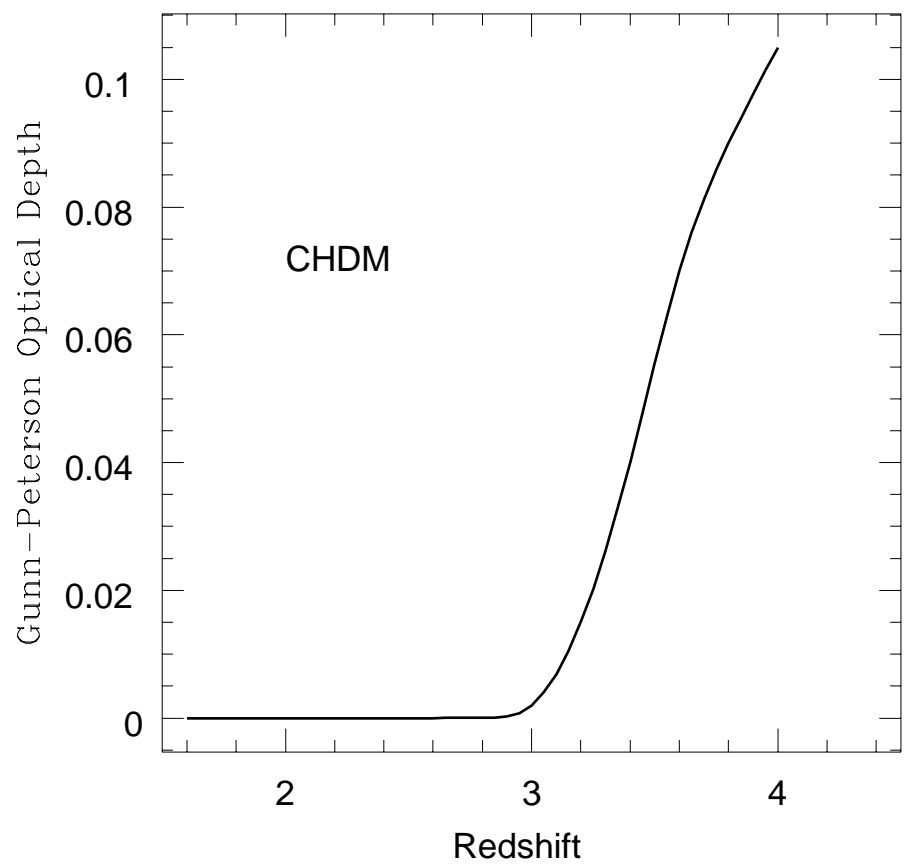

Fig. 12

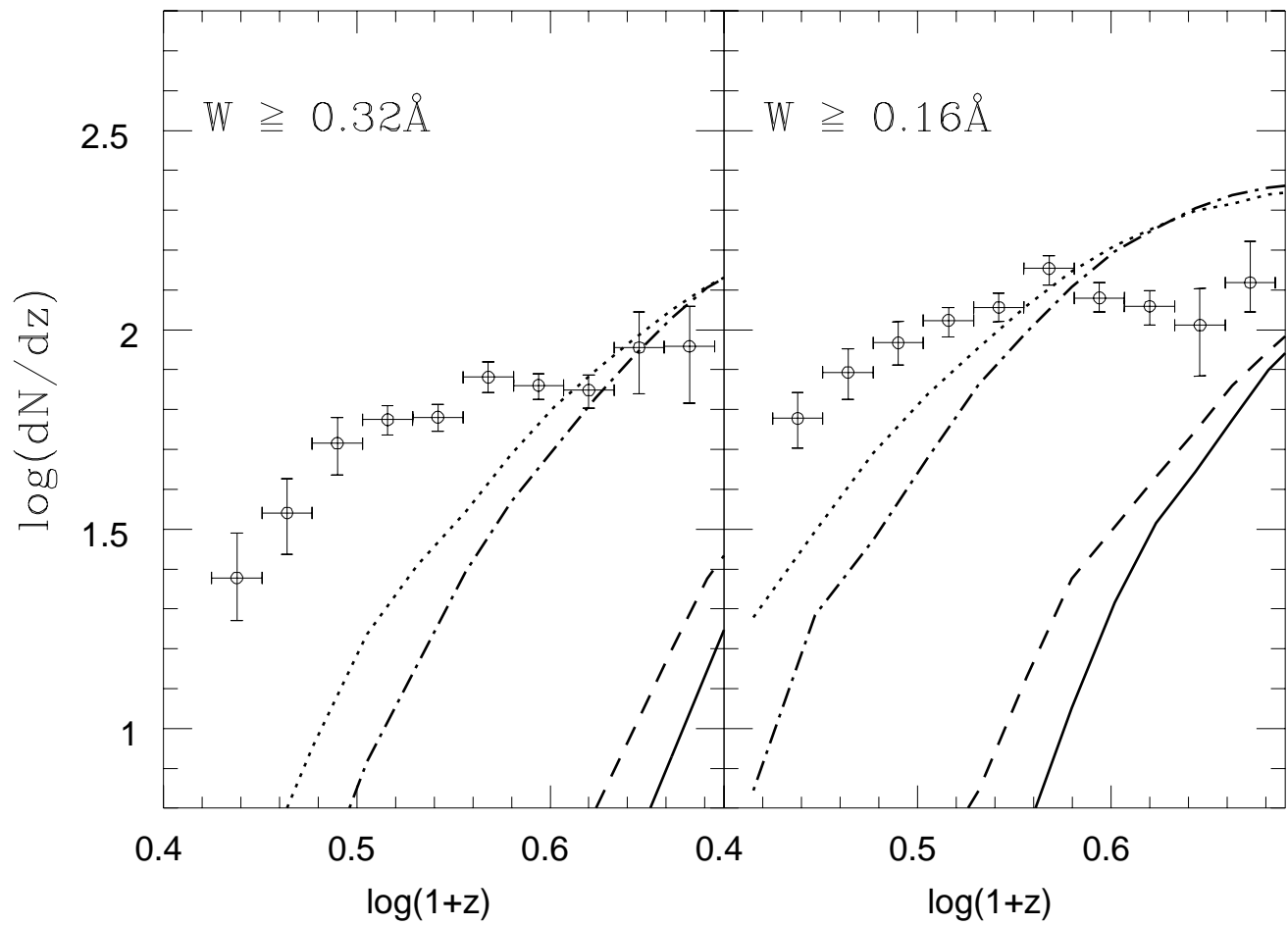

Fig. 13

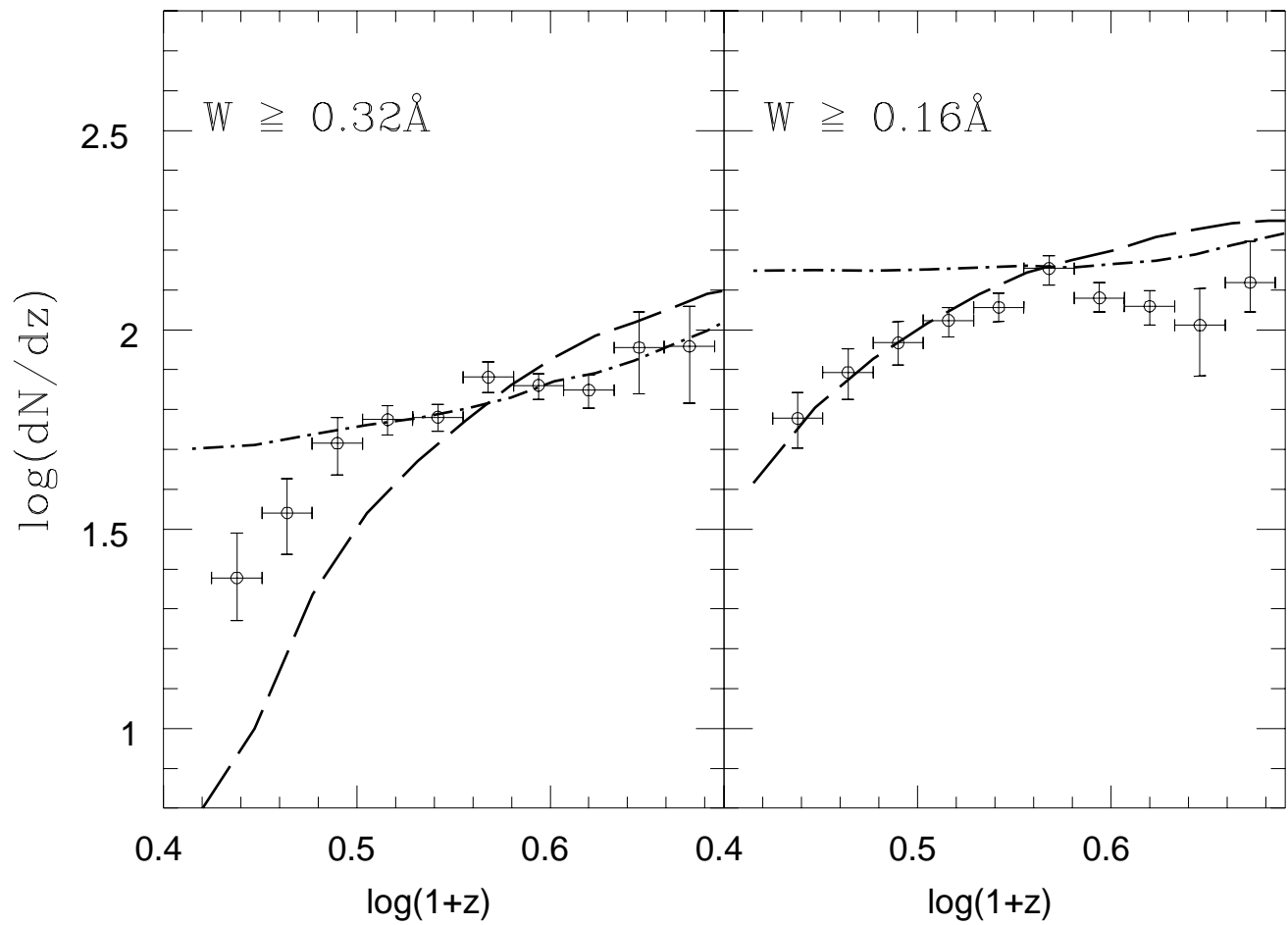

Fig. 14